\documentclass[12pt]{article}
\usepackage{epsfig,cite}
\newcommand{\non}{\nonumber}
\newcommand{\bea}{\begin{eqnarray}}
\newcommand{\eea}{\end{eqnarray}}
\newcommand{\bdm}{\begin{displaymath}}
\newcommand{\edm}{\end{displaymath}}

\def\be{\begin{equation}}
\def\ee{\end{equation}}

\def\hlf{\frac{1}{2}}

\def\drbar{\overline{\rm DR}}
\def\msbar{\overline{\rm MS}}
\def\sq2{\sqrt{2}}

\newcommand{\as}{\alpha_s}
\def\bsg{B\to X_{\!s}\, \gamma}

\newcommand{\smallw}{{\scriptscriptstyle W}} %

\newcommand{\muw}{\mu_\smallw}
\newcommand{\susy}{\rm{\scriptscriptstyle SUSY}}
\newcommand{\mumfv}{\mu_{{\scriptscriptstyle MFV}}}
\newcommand{\mususy}{\mu_{\susy}}
\newcommand{\msusy}{M_{\susy}}

\def\mgl{m_{\tilde{g}}}
\def\mhc{m_{H^\pm}}

\def\msbi{m_{\tilde{b}_i}^2}

\def\sbl{\tilde{b}_L}
\def\sbr{\tilde{b}_R}
\def\sblc{\tilde{b}^*_L}
\def\sbrc{\tilde{b}^*_R}
\def\sbu{\tilde{b}_1}
\def\sbd{\tilde{b}_2}

\def\sbi{\tilde{b}_i}

\def\mssl{m_{\tilde{s}_L}^2}

\def\ssl{\tilde{s}_L}
\def\ssr{\tilde{s}_R}
\def\sslc{\tilde{s}^*_L}
\def\ssrc{\tilde{s}^*_R}

\def\sdi{\tilde{d}_i}
\def\sdu{\tilde{d}_1}
\def\sdd{\tilde{d}_2}
\def\sdt{\tilde{d}_3}

\def\sdic{\tilde{d}^*_i}

\def\dul{d_{1\,L}}
\def\ddl{d_{2\,L}}

\begin{document}
\thispagestyle{empty}
\vspace*{-15mm}
 
\begin{flushright}
RM3-TH/06-1\\
DFTT-03/2006\\
LAPTH-1136/06\\

\vspace*{2mm}
\end{flushright}
\vspace*{20mm}
 
\boldmath
\begin{center}
{\LARGE{\bf
QCD Corrections to Radiative $B$ Decays in \\
\vspace*{3mm}  the MSSM with Minimal Flavor Violation
}}
\vspace*{5mm}

\end{center}
\unboldmath
\smallskip
\begin{center}
{\Large{G.~Degrassi$^a$, P.~Gambino$^b$, and P.~Slavich$^c$}} \vspace*{8mm} \\
{\sl ${}^a$
    Dipartimento di Fisica, Universit\`a di Roma Tre and  INFN, Sezione di
    Roma III, \\
    Via della Vasca Navale~84, I-00146 Rome, Italy}\\
\vspace*{2.5mm}
{\sl  ${}^b$ INFN, Sezione di Torino and Dipartimento di Fisica Teorica, 
Universit\`a di Torino,\\ Via P.~Giuria 1,  I-10125 Torino, Italy}  
\vspace*{2.5mm}\\
{\sl ${}^c$  LAPTH, 9, Chemin de Bellevue, F-74941 Annecy-le-Vieux,  France}
\vspace*{18mm}

{\bf Abstract}\vspace*{-.9mm}\\
\end{center}
\noindent

We compute the complete supersymmetric QCD corrections to the Wilson
coefficients of the magnetic and chromomagnetic operators, relevant in
the calculation of $b\to s \gamma$ decays, in the MSSM with Minimal
Flavor Violation. We investigate the numerical impact of the new
results for different choices of the MSSM parameters and of the scale
where the quark and squark mass matrices are assumed to be aligned. We
find that the corrections can be important when the superpartners are
relatively light, and that they depend sizeably on the scale of
alignment.  Finally, we discuss how our calculation can be employed
when the scale of alignment is far from the weak scale.

\setcounter{page}{0}
\vfill

\newpage
\section{Introduction}
More than a decade after their first direct observation, radiative $B$
decays have become a key element in the program of precision tests of
the Standard Model (SM) and its extensions.  The inclusive decay
$\bsg$ is particularly well suited to this precision program thanks to
its low sensitivity to non-perturbative effects.  The present
experimental world average \cite{hfag} for the branching ratio of
$\bsg$ has a total error of less than 10\% and agrees well with the SM
prediction, that is subject to a similar uncertainty \cite{GM}.  In
view of the final accuracy expected at the $B$ factories, about 5\%,
the SM calculation needs to be improved. It presently includes
next-to-leading order (NLO) perturbative QCD corrections as well as
the leading non-perturbative and electroweak effects (see
\cite{beauty} and \cite{Buras:2002er} for a concise discussion and a
complete list of references).  The calculation of
next-to-next-to-leading order (NNLO) QCD effects is currently under
way \cite{nnlo} and is expected to bring the theoretical accuracy to
the required level.

The theoretical accuracy of the predictions in the context of new
physics models may have important consequences on model building. This
is particularly true for radiative $B$ decays, where higher order
corrections can be enhanced by large factors: in this case the current
status of theoretical calculations is not always satisfactory. While
the NLO corrections have been extensively studied in the context of
Two Higgs Doublet Models \cite{CDGG1,2HDM,GM}, in the Minimal
Supersymmetric Standard Model (MSSM) the complete leading order (LO)
result is known \cite{LO,Borzumati:1999qt} but the NLO analysis is
still incomplete to date.  The main reason is that new sources of
flavor violation generally arise in the MSSM, making a general
analysis quite complicated even at the leading order
\cite{Borzumati:1999qt}.  Experimental constraints on generic $b\to s$
flavor violation have been recently studied in \cite{GFV}: radiative
decays play a central role in these analyses, and the constraints are
strong only for some of the flavor violating parameters.

A simplifying assumption frequently employed in supersymmetric
analyses is that of {\it Minimal Flavor Violation} (MFV), according to
which the only source of flavor (and possibly of CP) violation in the
MSSM is the CKM matrix \cite{DGIS,MFV}.  It can be implemented by
assuming that the squark and quark mass matrices can be simultaneously
diagonalized and, as a consequence, it implies the absence of
tree-level flavor-changing gluino (FCG) interactions.  The MFV
hypothesis certainly represents a useful and predictive approximation
scheme and seems to be favored by the present absence of deviations
from the SM. However, because the weak interactions affect the squark
and quark mass matrices in a different way \cite{wyler}, their
simultaneous diagonalization is not preserved by higher order
corrections and can be consistently imposed only at a certain scale
$\mumfv$, complicating the study of higher order contributions in this
framework.  The NLO study of radiative decays in the MFV scenario has
been pioneered in \cite{CDGG2} (see also \cite{bobeth}), where the
gluonic corrections to chargino contributions have been computed,
while those involving a gluino were computed in the heavy gluino
limit, in which case FCG effects can be consistently neglected.

An alternative possibility is to include only the potentially 
large contributions beyond the leading order: they originate from terms
enhanced by $\tan\beta$ factors, when the ratio between the two Higgs
vacuum expectation values is large, or by logarithms of $\msusy/M_W$,
when the supersymmetric particles are considerably heavier than the
$W$ boson.  Compact formulae that include both kinds of higher-order
effects within MFV are given in ref.~\cite{DGG}.  Indeed,
$\tan\beta$-enhanced terms at the next-to-leading order do not only
appear from the Hall-Rattazzi-Sarid effect (the modified relation
between the bottom mass and Yukawa coupling) \cite{HRS}, but also from
an analogous effect in the top-quark Yukawa coupling \cite{DGG,carena,
DGIS} and in effective flavor-changing $\bar{s}_L b_R$ neutral heavy
Higgs vertices \cite {DGIS}.  In the effective theory approach first
employed in \cite{carena1} the dominant terms enhanced by $\tan\beta$
can be taken into account at all orders. A generalization beyond MFV has 
been proposed in \cite{rosz2}.

In the limit of heavy superpartners, in particular, the Higgs sector
of the MSSM is modified by non-decoupling effects and can differ
substantially from the type-II Two Higgs Doublet Model.  The charged
Higgs contribution therefore receives two-loop contributions enhanced
by $\tan\beta$ that have been computed in \cite{DGG,carena,DGIS} in
the limit of heavy gluino. Interestingly, the explicit calculation of
the relevant two-loop diagrams presented in \cite{borzu} has
demonstrated the validity of this approximation even when the charged
Higgs is not much lighter than the gluino. However, there is a priori
no reason why the results derived in the heavy gluino limit should be
a good approximation of the true result for generic values of the
relevant mass parameters or in the case of other two-loop
contributions.

In this letter we present the results of the full NLO calculation of
the supersymmetric QCD corrections to the Wilson coefficients of the
two operators that are relevant in the MFV scenario, extending and
completing the work of ref.\cite{CDGG2}. In particular, we compute all
two-loop diagrams that contain a gluino, under the assumption that the
gluino couplings to quarks and squarks are flavor conserving at the
scale $\mumfv$.  Our results allow for a consistent and complete NLO
analysis of radiative $B$ decays in the MFV framework.

The paper is organized in the following way: in section 2 we describe
the calculation, the renormalization procedure, and the checks; in
section 3 the numerical impact of our results is discussed; section 4
explains how our results can be employed in the context of realistic
models of SUSY breaking and contains our conclusions.

\section{Gluino contribution to the Wilson coefficients}
As we focus here on short-distance contributions with MFV, we can
restrict our discussion to the form of the Wilson coefficients of the
$\Delta B=1$ magnetic and chromo-magnetic operators\footnote{ There
are one-loop gluino contributions to the Wilson coefficients of the
four-quark operators, but we will not consider them here.}
$Q_7=(e/16\pi^2)m_b {\bar s}_L \sigma^{\mu \nu}b_RF_{\mu\nu}$ and
$Q_8=(g_s/16\pi^2)\,m_b {\bar s}_L \sigma^{\mu \nu}T^ab_RG^a_{\mu\nu}$
evaluated at the matching scale $\muw$ in the effective Hamiltonian:
\be
\label{effH}
{\cal H}= -\frac{4G_F}{\sqrt{2}}V_{ts}^*V_{tb} \sum_i C_i(\muw) Q_i(\muw)
\ee
where $G_F$ is the Fermi constant and $V_{ts},\,V_{tb}$ are elements
of the CKM matrix.  We can organize the Wilson coefficients of the
operators $Q_{7,8}$ in the following way:
\bea
C_{7,8}(\muw)& = &C^{(0){\,\rm SM}}_{7,8}(\muw) + 
 C_{7,8}^{(0){\,H^\pm}}(\muw)
 + C_{7,8}^{(0){\,\susy}}(\muw)  \non\\ &+&
\frac{\as(\muw)}{4 \pi} \left[ C^{(1){\,\rm SM}}_{7,8}(\muw) +
 C_{7,8}^{(1){\,H^\pm}}(\muw)
 +  C_{7,8}^{(1){\,\susy}}(\muw) \right],
\label{wc1}
\eea
where the various LO contributions are classified according to whether
the corresponding one-loop diagrams contain only SM fields
($C_{7,8}^{(0)\,\rm SM}$), a physical charged Higgs boson and an up-type quark
($C_{7,8}^{(0)\,H^\pm}$), or a chargino and an up-type squark
($C_{7,8}^{{(0)\,\susy}}$).  The expressions for $C_{7,8}^{(0){\,\rm
SM}}$ and $C_{7,8}^{(0){\,H^\pm}}$ can be found, e.g., in
ref.~\cite{CDGG1}, while those for $C_{7,8}^{(0){\,\susy}}$ can be
found, e.g., in eq.~(4) of ref.~\cite{CDGG2}. Neutralino and gluino
exchange diagrams will be neglected under our MFV assumption.  The
relation between the LO and NLO Wilson coefficients at $\muw$ and the
branching ratio for $\bsg$ is well known (see for example
refs.~\cite{GM,CDGG1}).

\begin{figure}[p]
\begin{center}
\mbox{\epsfig{file=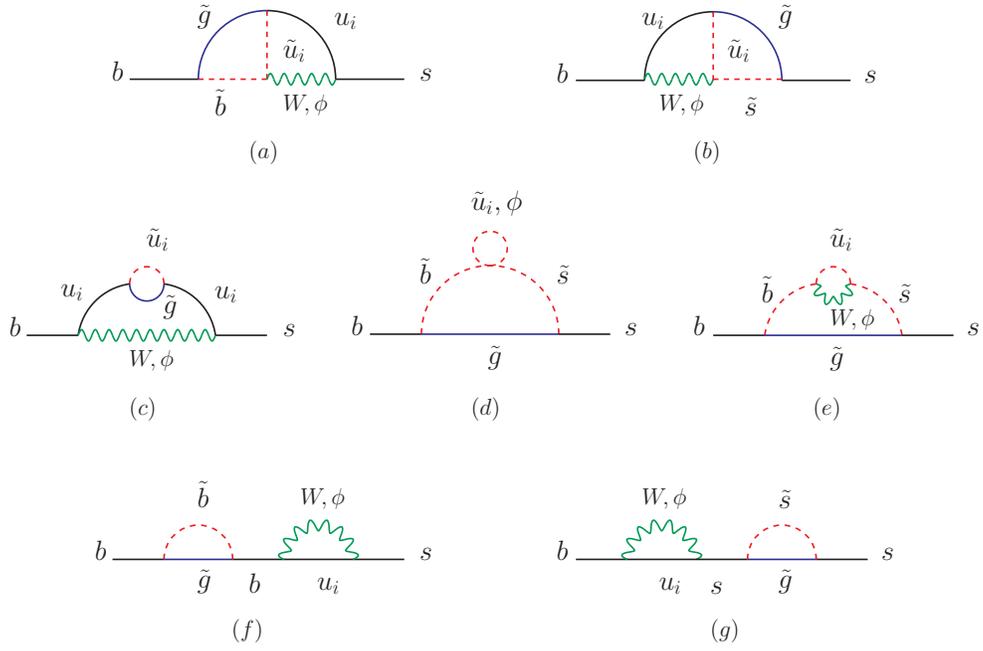,width=13.5cm}}
\end{center}
\vspace{-2mm}
\caption{\sf Feynman diagrams containing a gluino and a $W$ or a Higgs
boson ($\phi = H^\pm,G^\pm$).  A photon or gluon is assumed to attach
in all possible ways to the particles in the loops.} 
\label{diag1}
\end{figure}
\begin{figure}[p]
\begin{center}
\mbox{\epsfig{file=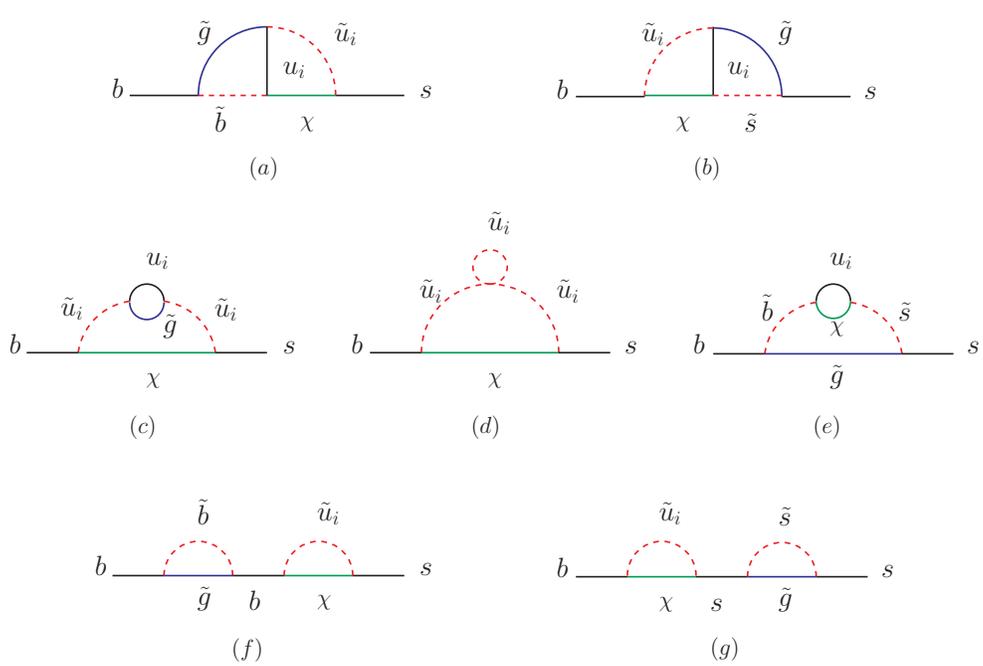,width=13.5cm}}
\end{center}
\vspace{-2mm}
\caption{\sf Same as fig.~1 for diagrams containing a chargino and a
gluino or a quartic squark coupling. The index $i$ labels the three
generations of up-type quarks and squarks.} 
\label{diag2}
\end{figure}

The NLO coefficients $C^{(1){\,\rm SM}}_{7,8}$ and
$C_{7,8}^{(1){\,H^\pm}}$ contain the gluonic two-loop corrections to
the SM and charged Higgs loops, respectively, and can be found for
instance in ref.~\cite{CDGG1}.  At NLO the supersymmetric contribution
$ C_{7,8}^{(1){\,\susy}}$ can be further decomposed,
\be
\label{decom}
C_{7,8}^{(1){\,\susy}} = C_{7,8}^{(1){\,\tilde{g}}} +
C_{7,8}^{(1){\,\chi^\pm}}, 
\ee 
where $C_{7,8}^{(1)\,\tilde{g}}$ contains two-loop diagrams with a
gluino together with a Higgs or W boson, while
$C_{7,8}^{(1){\,\chi^\pm}}$ corresponds to two-loop diagrams with a
chargino together with a gluon or a gluino or a quartic squark
coupling.  It should be recalled that, unlike $C^{(1){\,\rm
SM}}_{7,8}$ and $C_{7,8}^{(1){\,H^\pm}}$, the two-loop gluonic
corrections to the chargino loops are not UV finite: as shown in
\cite{CDGG2}, in order to obtain a finite result one has to combine
them with the chargino-gluino diagrams.  The chargino-gluon two-loop
contributions have been fully computed in refs.~\cite{CDGG2,bobeth}.
On the other hand, two-loop contributions involving gluinos (in both
$C_{7,8}^{(1){\,\tilde{g}}}$ and $C_{7,8}^{(1){\,\chi^\pm}}$) have
been considered in ref.~\cite{CDGG2} only in the heavy gluino
limit\footnote{In ref.~\cite{CDGG2} one of the top squarks was also
decoupled, but it is straightforward to generalize the formulae for
the light stop to the heavy stop.  }.  We are now going to relax this
approximation and to compute $ C_{7,8}^{(1){\,\susy}}$ for arbitrary
gluino mass in the MFV framework, assuming vanishing flavor-changing
gluino couplings.

The two-loop diagrams containing a gluino or a quartic squark coupling
that contribute to $C_{7,8}^{(1)\,\tilde{g}}$ and
$C_{7,8}^{(1)\,\chi^\pm}$ are shown in figs.~\ref{diag1} and
\ref{diag2}, respectively. Together with the diagrams with gluons,
they complete the QCD contribution to the Wilson coefficients of
$Q_{7,8}$ in the MSSM under the MFV assumption. The effective theory
is trivial, and the Wilson coefficients are directly given by the
result of the Feynman diagrams. We follow the same methods employed in
\cite{CDGG1}, in particular we perform our calculation in the
background-field gauge \cite{bground}, regularize the ultraviolet
divergences using naive dimensional regularization (NDR), and neglect
terms suppressed by powers of $m_b/M_W$ or $m_b/\msusy$ (after
factoring out a bottom mass in the definition of the operators $Q_7$
and $Q_8$). The result for each diagram depends on a number of mass
and coupling parameters; it can be simplified assuming the up-type
squarks of the first two generations to be degenerate in mass, and
neglecting the masses of all quarks of the first two generations. This
set of assumptions allows us to exploit the unitarity of the CKM
matrix and to factor out the combination $V_{ts}^*V_{tb}$ in the
effective Hamiltonian of eq.~(\ref{effH}).

The complete calculation of the two-loop gluino contribution presents
a novel feature with respect to heavy gluino analysis of \cite{CDGG2},
namely the need for flavor-changing counterterms.  Indeed, there are
two-loop gluino diagrams that contain the effective FCG interactions
$\tilde{b} \tilde{g} s$ or $b \tilde{s} \tilde{g}$ (see, e.g.,
diagrams ({\it a}) and ({\it b}) in fig.~\ref{diag1} and \ref{diag2},
respectively). These one-loop electroweak vertices are divergent and
need to be renormalized. The corresponding contributions were
irrelevant in \cite{CDGG2} because they are suppressed by inverse
powers of the gluino mass. We therefore distinguish between
flavor-conserving counterterms, already considered in \cite{CDGG2},
and flavor-changing counterterms of electroweak origin.

{\it Flavor-conserving} counterterms are of ${\cal O}(\as)$ and
originate from the masses of the bottom and top quarks, from the
masses and left-right mixing of the up-type squarks that enter the
one-loop diagrams with charginos, and from the flavor diagonal part of
the external leg corrections. The finite parts of these counterterms
depend on our choice of renormalization scheme for the masses and
mixing angles that enter the one-loop results.  In order to facilitate
the inclusion and resummation of some large higher order effects, one
can also distinguish between the top and bottom masses that originate
from the loops or from the use of equations of motion, and those
arising from Yukawa couplings or their supersymmetric equivalent.

In the MFV framework, the remaining {\it flavor-changing} counterterms
are of electroweak origin and arise from the renormalization of the
flavor mixing of quarks and squarks and from the flavor changing part
of the external leg corrections.  To discuss them, we start from the
gluino-quark-squark interaction Lagrangian in the super-CKM basis,
where the matrices of Yukawa couplings are diagonal and the squarks
are rotated parallel to their fermionic superpartners:
\be
\label{largorig}
{\cal L} \supset -  g_s \,T^a\, \sq2\,\left(
\bar{g^a}\, b_L \,\sblc - \bar{g^a}\, b_R \,\sbrc
+ \bar{g^a}\, s_L \,\sslc - \bar{g^a}\, s_R \,\ssrc 
\right) \,+ {\rm h.c.}
\ee
where $g_s$ is the strong coupling constant and $T^a$ are SU(3)
generators. We can restrict to the mixing between second and third
generations, and since we are neglecting $m_s$, we need not consider
the terms involving $s_R$ or $\ssr$.  Upon renormalization of the
mixing matrices, the {\em bare} quark and squark fields are rotated as
follows:
\be
\label{rotq}
\left(\begin{array}{c}\sdu\\ \sdd\\ \sdt\end{array}\right)
= \;(U^r + \delta U)\;
\left(\begin{array}{c}\sbl\\ \sbr\\ \ssl\end{array}\right)\,,
\;\;\;\;\;\;\;\;
\left(\begin{array}{c}\dul\\ \ddl \end{array}\right)
= \;(u^{L\,r} + \delta u^L)\,
\left(\begin{array}{c} b_L \\ s_L \end{array}\right)\,.
\ee 
The MFV assumption translates into the requirement that the
renormalized mixing matrices be flavor diagonal:
\be
\label{rotsq}
U^r = \left(\begin{array}{cc} 
B&0\\0&1\end{array}\right)
= \left(\begin{array}{ccc} 
\;\;\;\cos\theta_{\tilde{b}}&\sin\theta_{\tilde{b}}&0\\
-\sin\theta_{\tilde{b}}&\cos\theta_{\tilde{b}}&0\\0&0&1\end{array}\right)\,,
\;\;\;\;\;\;\;\;
u^{L\,r} = \left(\begin{array}{cc} 
1&0\\0&1\end{array}\right)\,,
\ee
where $B$ is a $2\times2$ mixing matrix in the sbottom sector and
$\theta_{\tilde{b}}$ is the sbottom mixing angle. Under this
requirement, the mass eigenstates for the down-type squarks relevant
to our calculation can be identified with the usual sbottoms $\sbu$
and $\sbd$ and the left super-strange $\ssl$.  However, even if we
assume that the renormalized mixing matrices for quarks and squarks
are flavor-diagonal, this is not the case for the corresponding
counterterms $\delta u^{L}$ and $\delta U$. They generate the FCG
interactions:
\be
\label{lagrot}
{\cal L} \supset - g_s \,T^a\, \sq2\,\left[\;
(\delta U^{\dagger}_{3i}+B^{\dagger}_{1i}\,\delta u^{L}_{21}) 
\,\overline{s_L}\, g^a\,\sbi 
+(\delta U_{31}-\delta u^{L}_{21}) \,\bar{g^a}\, b_L\,\sslc
-\delta U_{32} \,\bar{g^a}\, b_R\,\sslc 
\right] \,+ {\rm h.c.}
\ee
Additional flavor changing renormalization effects are due to the
(on-shell) wave function renormalization (WFR) of external quarks (see
diagrams ({\it f}) and ({\it g}) in figs.~1 and 2).  The divergent
parts of the mixing counterterms $\delta u^{L}$ and $\delta U$ are
determined in a gauge-invariant way by the requirement that they
cancel the divergence of the antihermitian part of the corresponding
WFR matrix \cite{mixing}.  Using $ m_s\to 0$ and neglecting terms
suppressed by $m_b/M_W$, we obtain:
\be
\label{countq}
\delta u^{L}_{21} = -\frac{1}{2}\,\left[ \,\Sigma^{L}_{sb}(0) + 2
 \,\Sigma^{S}_{sb}(0)\right]\,, \ee
where we have decomposed the generic quark self-energy as 
\be
\Sigma_{ij}(p) \;\equiv\; \Sigma^L_{ij}(p^2)\not{\!p}\,P_L 
+ \Sigma^R_{ij}(p^2)\not{\!p}\,P_R +  \Sigma^{S}_{ij}(p^2) (m_i P_L
+ m_j P_R)\,,
\ee
$P_L$ and $P_R$ being chiral projectors.  The counterterm for the
squark mixing matrix is instead 
\be
\label{defcountsq}
\delta U_{ik} = \hlf\,\Sigma_{j\neq i} \, 
\frac{ \Pi_{ij}(m_j^2)+\Pi^*_{ji}(m_i^2)}{m_i^2-m_j^2}\,U_{jk}\,,
\ee
which, for the terms that appear in eq.~(\ref{lagrot}), specializes to:
\be
\label{countsq}
\delta U^{\dagger}_{3i} =  \hlf\,
\frac{ \Pi_{\ssl\sbi}(\mssl)+\Pi_{\ssl\sbi}(\msbi)}{\msbi-\mssl}
\,,\;\;\;\;\;\;\;\;
\delta U_{3j} = -\hlf\,\Sigma_i \, 
\frac{ \Pi_{\ssl\sbi}(\mssl)+\Pi_{\ssl\sbi}(\msbi)}{\msbi-\mssl}\,B_{ij}\,.
\ee
We therefore see that the counterterms of the gluino flavor changing
couplings are determined by quark and squark flavor-changing two-point
functions only.  We have checked that the counterterms in
eqs.~(\ref{countq}) and (\ref{countsq}) renormalize correctly the
$\tilde{d}\,\bar{d'}\,g^a$ vertex (see \cite{Hikasa}) and agree with
the known one-loop RGE equations of the MSSM \cite{RGE}.

The finite part of the counterterms in eqs.~(\ref{countq}) and
(\ref{countsq}) is related to the way we interpret the MFV requirement
in eq.~(\ref{rotsq}).  In particular, if we perform a minimal
subtraction we are imposing the MFV condition on the
$\msbar$-renormalized parameters of the Lagrangian evaluated at the
scale $\mumfv$. An alternative option consists in absorbing also the
finite part of the antihermitian WFR: this results in a conventional
and gauge-dependent\footnote{ For the quark mixing matrix a
simplification occurs when the external quark masses can be neglected,
as in our case, and the gauge dependence drops out.} on-shell
renormalization scheme \cite{mixing}. In the following, we will assume
the first option and therefore our result will depend on the mass
scale $\mumfv$, that we identify with the scale where the quark and
squark mass matrices are assumed to be aligned.

Once the flavor-changing vertices of eq.~(\ref{lagrot}) are inserted
into one-loop diagrams with a gluino and a down-type squark, the
resulting counterterm contributions cancel the UV poles arising from
i) the diagrams in figs.~1$d$, 1$e$ and 2$e$, ii) the diagrams in
figs.~2$a$ and 2$b$ with the photon or gluon attached to the down-type
squark or to the gluino and iii) the flavor-changing WFR diagrams in
figs.~1$f$, 1$g$, 2$f$ and 2$g$. The remaining UV poles of the
diagrams in figs.~1 and 2 are canceled by the flavor-conserving
counterterms, but for a residual pole in the diagrams with gluino and
chargino of fig.~2. This is the pole that was found in \cite{CDGG2} in
the limit of heavy gluino; it is compensated by a corresponding pole
in the diagrams with gluon and chargino. In the gluonic corrections to
the chargino diagrams reported in \cite{CDGG2,bobeth} the residual UV
divergence has been subtracted either by the heavy gluino effective
chargino-quark-squark vertex or in a minimal way. The finite parts
related to this subtraction must be taken into account before
combining with the gluino contributions. A shift in the $\chi
\bar{b}\tilde{t}$ coupling is also necessary to restore supersymmetric
Ward identities that are not respected by NDR (see \cite{CDGG2}).

The analytic expressions of $C_{7,8}^{(1){\,\susy}}$ we derived are
too long to be reported. However, in view of our choice for the flavor
changing counterterms, we can split our result into two pieces
\be 
\label{peppe}
C_{7,8}^{(1){\,\susy}}(\muw) =
C_{7,8}^{(1a){\,\susy}}(\dots,\muw) + 
C_{7,8}^{(1b){\,\susy}}(\dots,\mumfv), 
\ee 
where the dots represent the relevant combination of couplings, masses
and mixing angles and the $(1a)$ piece can be identified with the
contribution that, in the heavy gluino limit, reduces to the result of
ref.\cite{CDGG2}. The interesting point is that $
C_{7,8}^{(1b){\,\susy}}$ contains logarithms of the ratio
$\msusy/\mumfv$, i.e. of a supersymmetric mass over a mass scale 
related to the mechanism of supersymmetry breaking. For example, in
supergravity models one identifies $\mumfv$ with the Planck mass and
therefore the Wilson coefficients contain very large logarithms that
need to be resummed.  If we were to employ an on-shell definition for
the flavor changing counterterms, $ C_{7,8}^{(1b){\,\susy}}$ would be
independent of $\mumfv$ and our result would have no large
logarithm. In practice, the use of on-shell mixing counterterms is
equivalent to assuming that MFV is valid at the scale of the
supersymmetric masses entering the loops.

We performed several checks of our calculation. Ref.~\cite{borzu}
presented a calculation of the $\tan\beta$-enhanced part of the
contribution to the Wilson coefficients coming from the diagrams in
fig.~1$b$ that involve a charged Higgs boson. We have verified that,
if we restrict our calculation to the same subset of diagrams and
adopt the same input parameters as in ref.~\cite{borzu}, we can
reproduce exactly fig.~8 of that paper. Also, a calculation of the QCD
contributions to the Wilson coefficients from the diagrams in
fig.~2$d$, involving a chargino and a quartic squark coupling, has
been presented in ref.~\cite{bobeth}. We have checked that, if we
assume MFV in the up squark sector and perform an $\msbar$
renormalization, we find complete agreement with the analytical
formulae of \cite{bobeth}. On the other hand, the contribution of the
diagrams in fig.~2$d$ is removed by the corresponding counterterm
contribution if the squark masses and mixing are defined on-shell.  As
already mentioned, the results for $C_{7,8}$ depend on the
renormalization scheme for a number of parameters. In the case all
parameters are renormalized in the on-shell scheme, the QCD
corrections to the Wilson coefficients still depend on the matching
scale $\muw$ at which the effective operators $Q_{7,8}$ are
renormalized (see eq.(\ref{effH})). This dependence can be expressed
in terms of the LO anomalous dimension matrix \cite{CDGG2} and we
reproduce it correctly.

\section{Numerical Results}

We start the discussion of our numerical results by defining the set
of input parameters relevant to the calculation of the Wilson
coefficients.  For the SM parameters we take $M_Z$ = 91.2 GeV,
$\sin^2\theta_W = 0.23$ and $\as(M_W) = 0.12$ and for the top mass we
use the SM value in the $\msbar$ scheme i.e.  $\overline{m}_t(M_W)$ =
176.5 GeV (corresponding to a physical top mass of 175 GeV).  The soft
SUSY-breaking terms that enter the squark mass matrices in the MFV
scenario and are relevant to our calculation are: the masses for the
SU(2) doublets, $m_{Q_i}$, where $i$ is a generation index; the masses
for the third-generation singlets, $m_T$ and $m_B$; the trilinear
interaction terms for the third-generation squarks, $A_t$ and
$A_b$. We take all of them as running parameters, computed in a
minimal subtraction scheme at the renormalization scale $\mususy =
500$ GeV.
We recall that, in the super-CKM basis, the $3\times3$ mass matrices
for the up-type and down-type left squarks are related by
$(M_U^2)_{LL} = V (M_D^2)_{LL} V^{\dagger}$, where $V$ is the CKM
matrix, therefore the two matrices can be both flavor-diagonal at
$\mu=\mumfv$ only if they are flavor-degenerate. This means that in
the MFV scenario we must introduce a common mass parameter for the
three generations of SU(2) squark doublets, i.e.~$m_{Q_i} \equiv m_Q$
at $\mu=\mumfv$.
The other MSSM parameters relevant to our calculation, for which we
need not specify a renormalization prescription, are: the charged
Higgs boson mass $\mhc$; the gluino mass $\mgl$; the SU(2) gaugino
mass parameter $M_2$; the higgsino mass parameter $\mu$, with the same
sign convention as in ref.~\cite{DGG}; the ratio of Higgs vacuum
expectation values $\tan\beta$.

Some potentially large higher-order corrections can be absorbed in the
one-loop results. Following ref.~\cite{DGG}, we absorb in the one-loop
coefficients the $\tan\beta$-enhanced corrections to the bottom Yukawa
coupling \cite{HRS}.  As explained in \cite{DGG}, large logarithms of
the ratio $\msusy/\muw$, induced by gluonic corrections to the
one-loop chargino-stop diagrams, could also be resummed to all orders
by expressing the higgsino couplings in terms of
$\overline{m}_t(\mususy)$. In what follows, however, we will use
$\overline{m}_t(M_W)$ for the couplings of charginos (or Higgs bosons)
to top quarks and squarks, as well as for the mass of the virtual top
quarks in the loops. For consistency with our choice of the SUSY
parameters, we will use in the stop mass matrix the
$\drbar$-renormalized top quark mass, computed at the scale $\mususy$
with the field content of the MSSM.

Leaving a systematic study of the constraints imposed on the MSSM
parameters by the $\bsg$ branching ratio to a future publication, we
restrict our analysis to two different choices of MSSM parameters:
\begin{itemize}

\item[(I)] $m_Q = 230$ GeV, $m_T = 210$ GeV,
$m_B = 260$ GeV, $A_t = -70$ GeV, $A_b = 0$, $\mhc= 350$ GeV,  $\mgl = M_2 =
200$ GeV, $\mu = 250$ GeV,  $\tan\beta = 30$;

\item[(II)] $m_Q = 480$ GeV, $m_T = 390$ GeV,
$m_B = 510$ GeV, $A_t = -560$ GeV, $A_b = -960$, 
$\mhc= 430$ GeV,  $\mgl = 600$ GeV, $M_2 = 190$ GeV, 
$\mu = 390$ GeV,  $\tan\beta = 10$.

\end{itemize}

The first set is analogous to ``spectrum II'' in ref.~\cite{borzu} and
is characterized by moderately large $\tan\beta$ and fairly light
superpartners. The second set of parameters corresponds broadly to the
so-called Snowmass Point SPS1a$^\prime$ \cite{snowmass}, obtained
through RG evolution from a set of universal high-energy boundary
conditions imposed by the mechanism of gravity-mediated supersymmetry
breaking. It is characterized by a smaller value of $\tan\beta$ and
somewhat heavier superpartners (well within the reach of future
collider experiments). In both cases we impose the MFV relation
$m_{Q_i} \equiv m_Q$.

\begin{figure}[p]
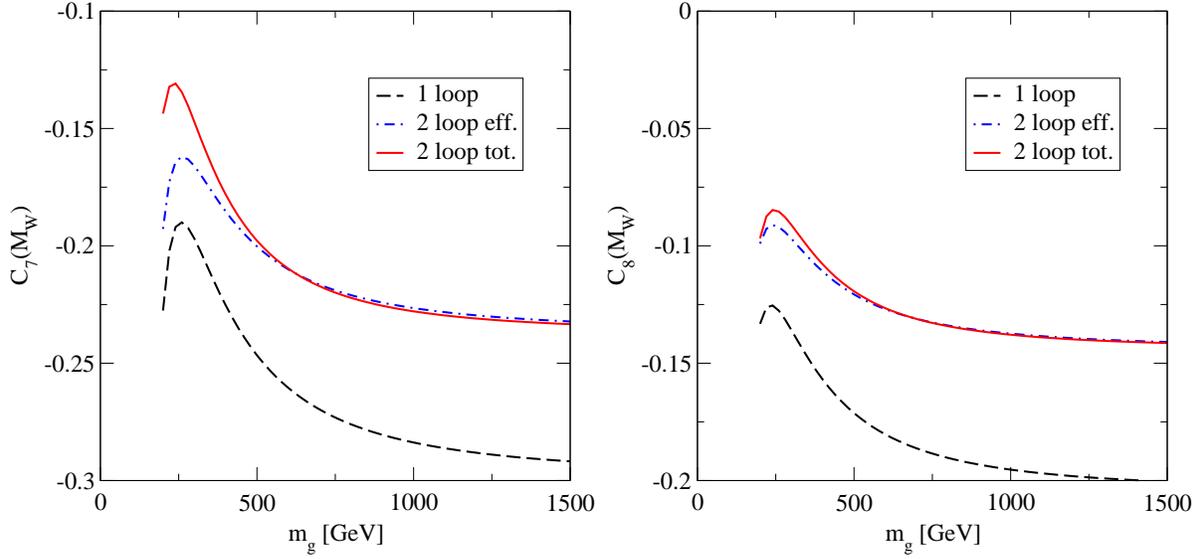

\begin{center}
\mbox{
\epsfig{figure=c7vsmg_I.eps,width=7.8cm}
\epsfig{figure=c8vsmg_I.eps,width=7.8cm}}
\end{center}
\caption{\sf Wilson coefficients $C_7(M_W)$, left, and $C_8(M_W)$, right,
as a function of the gluino mass for a choice of MSSM input parameters
modeled on set I (see text).  
}
\label{plot-I}
\end{figure}
\begin{figure}[p]
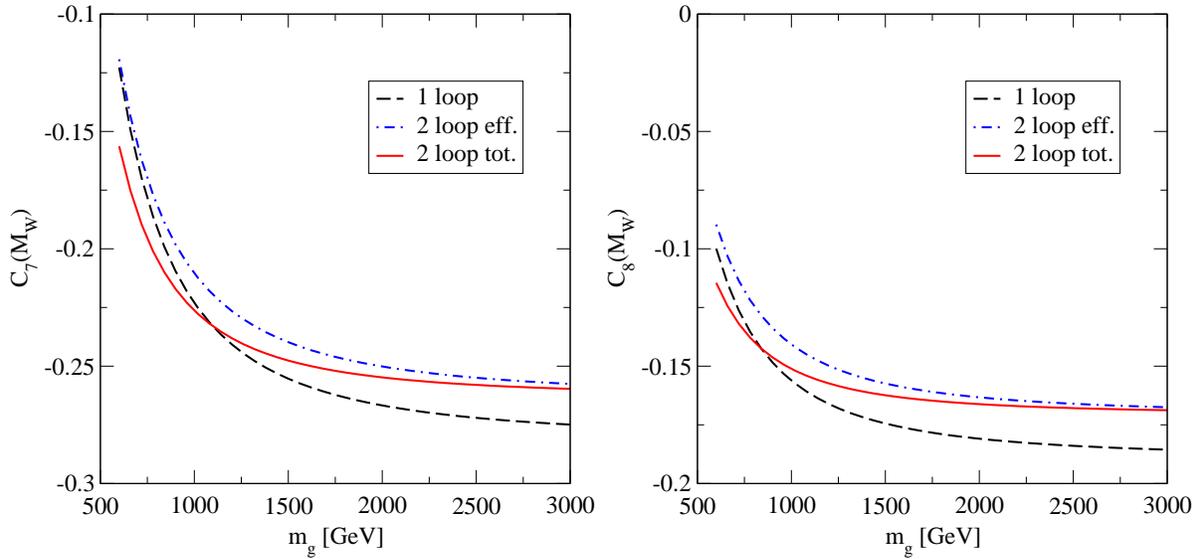

\begin{center}
\mbox{
\epsfig{figure=c7vsmg_IIp.eps,width=7.8cm}
\epsfig{figure=c8vsmg_IIp.eps,width=7.8cm}}
\end{center}
\caption{\sf Same as fig.~\ref{plot-I} for the MSSM input parameters
of set II.}
\label{plot-II}
\end{figure}

We can now discuss our numerical results for the Wilson coefficients
$C_7(M_W)$ and $C_8(M_W)$. To start with, we assume MFV at the level
of the running parameters of the MSSM Lagrangian, at the
renormalization scale $\mumfv=500$ GeV.  In figs.~\ref{plot-I} and
\ref{plot-II}, the left end of each curve corresponds to the choice of
MSSM input parameters defined above in the sets I and II,
respectively.  To study the decoupling behaviour of the corrections,
we rescale all the supersymmetric mass parameters -- but for the
charged Higgs boson mass -- by an increasing common factor, and show
$C_7(M_W)$ and $C_8(M_W)$ as a function of the resulting value of the
gluino mass (i.e.\ in figs.~\ref{plot-I} and \ref{plot-II} all the
squark and chargino masses increase together with $\mgl$). In each
plot, the dashed line corresponds to the pure one-loop result
(i.e. without resummation of the $\tan\beta$-enhanced corrections to
the bottom Yukawa coupling), supplemented with the two-loop gluonic
corrections to the diagrams with SM particles or charged Higgs boson;
the dot-dashed line contains in addition the $\tan\beta$-enhanced
gluino contributions as computed in the effective theory approach in
refs.~\cite{DGG,carena}; finally, the solid line corresponds to our
complete two-loop diagrammatic calculation.
Comparing the solid and dot-dashed curves in figs.~\ref{plot-I} and
\ref{plot-II}, it can be seen that for low values of the superparticle
masses the $\tan\beta$-enhanced gluino contributions of
refs.~\cite{DGG,carena} do not provide a good approximation of our
full two-loop result, especially in the case of $C_7(M_W)$. As the
superpartners get heavier, however, the effective theory approach
becomes more reliable, and the corresponding results get closer to
those of the complete calculation. Indeed, for large values of the
superparticle masses the difference between the two-loop results
(solid and dot-dashed lines) and the one-loop results (dashed lines)
is mainly due to the non-decoupling charged Higgs contributions
discussed in refs.~\cite{DGG,carena}.

\begin{figure}[t]
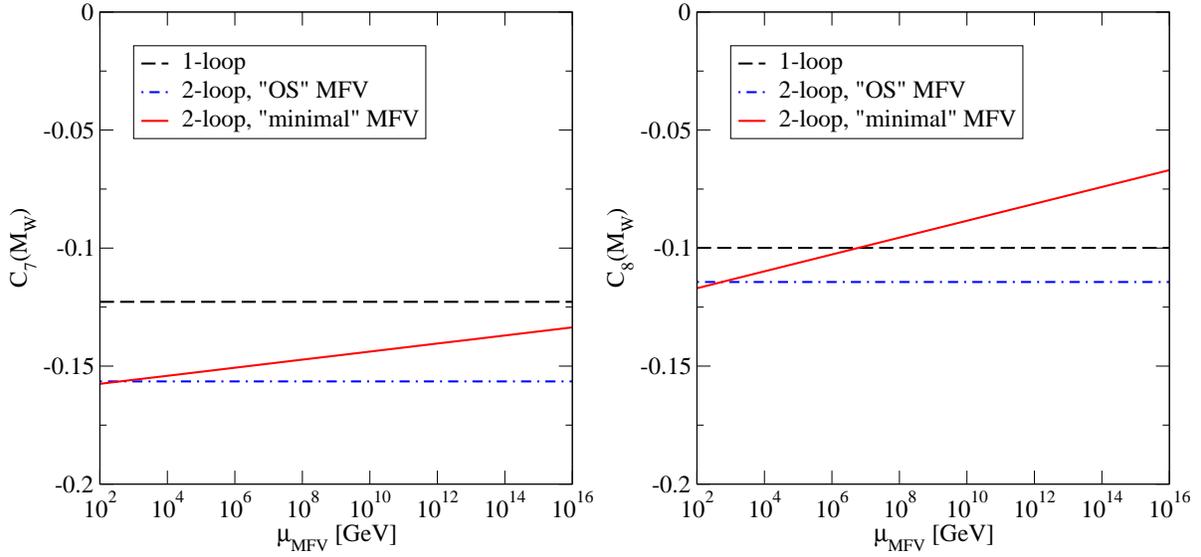

\begin{center}
\mbox{
\epsfig{figure=c7vsqmfv_IIp.eps,width=7.8cm}
\epsfig{figure=c8vsqmfv_IIp.eps,width=7.8cm}}
\end{center}
\vspace{-2mm}
\caption{\sf Wilson coefficients $C_7(M_W)$, left, and $C_8(M_W)$,
right, as a function of the scale $\mumfv$ at which the MFV condition
is imposed (see text).  }
\label{plot-mfv}
\end{figure}

As mentioned above, in figs.~\ref{plot-I} and \ref{plot-II} we assume
that MFV is valid at the level of the running parameters of the
Lagrangian, at a renormalization scale of the order of the
superparticle masses. The plots in fig.~\ref{plot-mfv}, obtained with
the MSSM parameters of set II, allow us to appreciate the implications
of this assumption. In each plot, the solid line represents our
two-loop results for the Wilson coefficients as a function of
$\mumfv$, when the latter is varied between 100 GeV and $10^{16}$
GeV. For comparison, we also plot the one-loop results (dashed lines),
defined as in the previous figures, and the scale-independent two-loop
results that we obtain by employing an on-shell definition of the
flavor changing counterterms (dot-dashed lines).
It can be seen that, for values of $\mumfv$ of the order of the
superparticle masses, the results obtained with the minimal definition
of the flavor changing counterterms are very similar to those obtained
with the on-shell definition. However, when $\mumfv$ is increased up
to the GUT scale, the logarithm of the ratio $\msusy/\mumfv$ becomes
very large, and the corresponding contribution modifies sensibly the
two-loop part of the correction.  Of course, in this case a
fixed-order calculation does not provide a good approximation to the
correct result, and the large logarithmic corrections have to be
resummed.

\section{Discussion and summary}
We have presented a complete calculation of the ${\cal O}(\as)$
supersymmetric corrections to the Wilson coefficients relevant for
radiative $B$ decays, assuming MFV (i.e.\ the vanishing of
flavor-changing gluino couplings) at a scale $\mumfv$.  The magnitude
of $\mumfv$ depends on the specific model of supersymmetry breaking,
but can be much larger than that of all other mass scales entering the
calculation, giving rise to large logarithms that must be resummed. It
is important to realize that the logs of $\mumfv$ are directly related
to the running of the flavor-changing gluino-quark-squark couplings
that we have required to vanish at that scale.  In other words, even
if we impose MFV at $\mumfv$, the MSSM lagrangian at a scale $\mu \neq
\mumfv$ will contain the interactions
\be
\label{lagrot2}
 - g_s \,T^a\, \sq2\,\left[\; g^i_{s_L}(\mu)
\,\overline{s_L}\, g^a\,\sdi + g^i_{b_L}(\mu)\,\bar{g^a}\,
b_L\,\sdic +g^i_{b_R}(\mu) \,\bar{g^a}\, b_R\,\sdic \right] \,+ {\rm
h.c.}, 
\ee 
where $\sdi$ are the down-type squark mass eigenstates (no longer
identified with flavor eigenstates). The couplings $g^i_{s_L}$ and
$g^i_{b_{R,L}}$ induce $b\to s$ transitions mediated by one-loop
gluino diagrams and their evolution follows from the standard RGE of
the MSSM (see \cite{wyler,RGE}). In particular, the resummation of the
large logs of $\mumfv$ is accomplished by solving the one-loop RGE for
the quark and squark mass matrices, which are then diagonalized at the
scale $\mu$. Indeed, the coefficient of $\log\mumfv$ in
eq.~(\ref{peppe}) can be easily reproduced by expanding the RGE
solution for the above couplings in powers of $\alpha_W$.

Even in the case of very large $\mumfv$, a natural and consistent
approximation scheme can be adopted if the $b\to s$ flavor violation
generated radiatively at the low scale $\mususy$, though not vanishing, is
{\em small} (as is generally the case for $\tan\beta$ not too large
\cite{borz}) or the gluino mass is large.  The one-loop gluino
diagrams can then be computed using the interactions in eq.~(13)
at the scale $\mususy$, and it is safe to neglect all QCD corrections
to this contribution. The same applies to one-loop diagrams with
flavor-changing neutralino-quark-squark couplings (whose contribution
gets also suppressed in the $Q_7\,$--$\,Q_8$ mixing \cite{DFMS}). In
addition to these two contributions, we are now able to include all
other supersymmetric contributions at ${\cal O}(\as)$. The QCD
correction $C_{7,8}^{(1){\,\susy}}(\muw)$, in particular, should be
computed using $\mumfv=\mususy$, because the radiative effects that
generate FCG interactions are already taken into account and resummed
by the one-loop gluino diagrams.  This strategy allows for a precise
calculation of radiative decays in the scenarios characterized by MFV
at a high scale. A detailed numerical implementation for the main SUSY
breaking scenarios will be presented elsewhere.

In summary, we have completed the calculation of the QCD corrections
to radiative $B$ decays in supersymmetric models characterized by
Minimal Flavor Violation at a scale $\mumfv$. In the case $\mumfv$ is
much larger than the electroweak scale, we have explained how to resum
the ensuing large logs. We have seen that the numerical results based
on the new calculation differ significantly from existing partial
calculations for relatively light superpartners, though they agree
well with \cite{DGG} in the case of a heavy SUSY spectrum.  We believe
the new results, soon to be made available as a public computer code,
will prove essential for an accurate calculation of radiative $B$
decays in most supersymmetric scenarios.

\section*{Acknowledgements}
We are grateful to Gian Giudice for collaboration at the early stage
of this project and for many useful discussions.  P.~S.\ carried out
part of this work at Durham IPPP, and is grateful to INFN, Torino for
hospitality. The work of P.~G.\ was supported in part by the EU grant
MERG-CT-2004-511156 and by MIUR under contract 2004021808-009. The
Feynman diagrams in figs.~1 and 2 were drawn with {\tt JaxoDraw}
\cite{jaxodraw}, and {\tt FormCalc} \cite{formcalc} was used to
convert our results into fortran routines.

\end{document}